\documentclass[aip,amsmath,amssymb,showpacs]{revtex4-1}
\usepackage{graphicx}
\usepackage{bm}

\begin{document}
\preprint{AIP/123-QED}
\topmargin -1.2cm
\topskip 15mm
\oddsidemargin -.2cm

\title{Energy of mixing and entropy of mixing for Cu$_{x}$Al$_{1-x}$ liquid binary alloys}
\author{Fysol Ibna Abbas}
\author{G. M. Bhuiyan}
\email{gbhuiyan@du.ac.bd}
\author{A.Z. Ziauddin Ahmed}    
\affiliation{Department of Theoretical Physics, University of Dhaka, Dhaka-1000, Bangladesh}
\begin{abstract}
The free energy of mixing and the entropy of mixing for Cu$_{x}$Al$_{1-x}$ liquid binary 
alloys have been systematically investigated by using the electronic theory of metals along with
the perturbation approach at a thermodynamic
state $T=1373$ K. The interionic interaction and a reference liquid are the fundamental components
of the theory. The interionic interaction is described by a local pseudopotential. A liquid of hard 
spheres (HS) of two different effective diametres is used to describe the reference system for alloys.
The results of the calculations for energy of mixing agree well with the available experimental data.
Calculation of entropy of mixing is parameter free and, the agreement with experiment, in this case,
is found to be fairly good.\\ 
\end{abstract}
\maketitle

\section{Introduction}
\noindent
In this article we have systematically investigated the energy of mixing and entropy of mixing for 
Cu$_{x}$Al$_{1-x}$ liquid binary alloys at temperature $1373$ K, from the first principles approach,
specifically from the perturbation method, and the electronic theory of metals (ETM). Recently, an
increasing  interest on the study of physical properties of Cu$_{x}$Al$_{1-x}$ binary alloys has been observed
\cite{1,2,3,4,5,6}. Study of mixing behavior of an alloy formed by two or more elemental metals is however important
not only to understand the above thermodynamic behaviors alone but also to understand underlying 
mechanisms of other characterisric properties of matter such as segragation of alloys, phase 
transition, the 
glass transition temperature, and stability of liquid metals and their alloys at different 
thermodynamic states. It is worth mentioning here that, the first application of the thermodynamic 
perturbation theory along with the Gibbs-Bogoliubov variational scheme for the liquid binary alloys
was due Umar $et$ $al.$  \cite{7}. Along the same line, thermodynamic properties are studied by some
authors \cite{8,9,10,11} and, thus advanched our knowledge. But in the present investigation we have used the first
principles perturbation approach without variational scheme \cite{12,13,14}. This approach uses the full potential profile 
in the theory and, allows us to understand different mechanisms involved in determining the mixing behavior
 of alloys. We further note that the present approach contains an important additional term in the theory
 than the approach of perturbation theory with variation\cite{7,8} provides.

The energy of mixing, and entropy of mixing are not independent to each other in thermodynamics, 
these are rather related by the following thermodynamic relation, 
\begin{equation}
\Delta G = \Delta H- T\Delta S 
\end{equation}
where $\Delta G$, $\Delta H$ denote the Gibbs free energy of mixing and enthalpy of mixing. We note that
at low pressure (one or two atmospheric pressure)  $\Delta G \simeq \Delta F$ and $\Delta H 
\simeq \Delta E$, where $\Delta F$ and 
$\Delta E$ are Helmholtz free energy of mixing and internal energy of mixing, respectively. We further note that,
at zero pressure the 
above relations between G and F, and H and E become exactly equal. Henceforth by the term energy of mixing we shall mean $\Delta F$
if not otherwise stated. However, evaluation of energy of mixing directly involves the full profile of 
the interionic pair potentials wheares the entropy of mixing is related directly to the derivative of
free energy with respect of temperature. So, it is always a challenging task to calculate 
entropy of mixing of liquid binary alloys theoretically. Because, the success of the entropy calculation
depends not on the free energy value but on the accuracy of the detailed shape of the free energy as a 
function of temperature. For example, 
Asta $et$ $al.$ \cite{15} shows that computer simulation along with different sophisticated EAM potentials 
\cite{16,17,18} can  predict the enthalpy
of mixing well, but fails to describe entropy of mixing on the same footing. It is worth nothing that the
energy of mixing predicts stability from the principle of minimization of free energy and, the entropy of mixing from
the principle of maximum entropy. But $\Delta F$ and $\Delta S$ together can put the prediction on the firm
basis. On the other hand, the knowledge of entropy may be used to study the atomic transport properties 
through different scaling laws \cite{19,20,21,22}.

Cu$_{x}$Al$_{1-x}$ binary alloy is formed by two different elements Al and Cu. Al in the solid phase has $fcc$
crystalline structure. It is a relatively soft, durable, light weight, ductile and malleable metal. Al is a good 
 electrical and thermal conductor, and also a superconductor. It is a polyvalent system with chemical valence of $3$. 
Elemental Cu has fcc structure in the solid phase and characterized by high ductility, electrical and thermal 
conductivity. Although $d$-shell in this elements is filled by electrons, the $d$-band effect to the interionic 
interactions through $s-d$ hybridization is still there, which sometimes plays significant role in determining 
physical properties of Cu. However, Cu$_{x}$Al$_{1-x}$ alloys display negative deviation from the Rault's law. This alloy
is characterized by a large amount of intermetallic phases in the solid state.

As discussed above, Al is a simple polyvalent metal
but Cu exhibits $d$-band effects thorugh the $sp$-$d$ hybridization. It is therefore urging a model for interionic 
interactions that can handle separately both simple metals with extended $sp$-band and transition metals 
with narrow $d$-band adequately. In this regard the 
Bretonnet-Silbert (BS) pseudopotential model \cite{23} is a promosing candidate, because, it treats $sp$ and $d$-band separately within the well established  pseudopotential theory. Here, the $sp$-band is described via the empty core model and, the $d$-band effect is derived from the $d$-band scattering phase shift by using an inverse scattering approach. The resulting model reduces to a simple local form which appears similar to that of well known Heine-Aberenkov (HA) model \cite{24}. The main difference between these two models is that the core term in the BS model is derived from the scattering phase shift and is a function of interionic distance and, in the HA model is a square well potential, which is derived by the quantum defect method. Moreover, the BS model has already been proven to be successful in the study of static structure of liquid metals and their alloys \cite{25,26,27,28}, atomic transport properties \cite{29,30,31}, thermodynamics 
\cite{32,33,34,35}, electronic transport properties  \cite{36} and even a complicated phenomena of segregation \cite{12,13,14}. Note that the norm-conserving non-local pseudopotentials \cite{40} are in principle, preferred for accurate predictions, but there are evidences that, the local potentials \cite{41} describe physical properties, in some cases, even better than the former ones.

In order to study the thermodynamics of mixing of liquid binary alloys several theories have been put 
forward; of them the quasilattice theory \cite{42}, the electronic theory of metals (ETM) \cite{7,9,34,35}, computer simulation method \cite{43}, empirical linear free energy \cite{44} and the free volume model \cite{1} are commonly used. In the semi-empirical quasi lattice theory the activity is expressed in terms of average interionic
 interaction energy and the formation of mixing via the Gibbs free energy. The enrgy of mixing is then evaluated by fitting to the experimental data of the activity. The linear free energy theory and the free volume model are totally empirical ones. The electronic theory of models (ETM) used in this work is based on the energy band structure of electons,
a characristic feature of metals, and thus evolved as a microscopic theory of metal. The static structure factor is determined from the knowledge of interionic pair interions, obtained from the band structure energy, through the statistical mechanics \cite{45}. Most importantly, in the present approach, each term in the ETM and the mechanism of its origin are clearly understandable from the physical point of view.

It is well known that the first principles perturbation theory requires such a reference system which can closely resemble the concerned  real system \cite{46}. There are many experimental as well as theoretical evidences that the HS theory of liquid can describe the structure of simple and transition metals and their binary alloys \cite{47,48,49}. Being prompted by the above history of success we employ HS reference system within the Percus-Yevick approximation (HSPY) \cite{50} in the present work. It is very much relevant to note here that, calculation of the static structure of HSPY liquid system requires the knowledge of effective hard sphere diameter (HSD). There are different methods to determine HSD for a liquid metal. Of them we have chosen the linearized Week-Chandler-Anderson (LWCA) termodynamic perturbation theory. The base of the LWCA theory is the WCA theory, which was first applied in Ref \cite{51}.

This report is organized in the following way. Theories relavant to the present calculations for thermodynamics of mixing are briefly described in section II. Section III is devoted to the presentation of results of calculation, and for discussion. We, finally, conclude this article with some remarks in section IV.
\section{Theory}
For calculating the energy of mixing and entropy of mixing for Cu$_{x}$Al$_{1-x}$ liquid 
binary alloy at $1373$ K we have used different theories. Some of the relevant theories namely the effective 
partial pair potentials, partial pair correlation functions for HS binary liquids, energy
of mixing and entropy of mixing for liquid binary alloy are briefly described below. 
\subsection{The effective partial pair potential}
The local pseudopotential for one component metallic systems
may be modeled by the superposition of the $sp$-band and $d$-band
contributions as \cite{23}, 
\begin{equation}
 V(r) =\begin{cases}
\sum_{m=1}^{2}B_{m}\exp(-r/ma) & \text{if  $r<R_{c}$}  \\
 -Z/r & \text{if $r > R_{c}$}
\end{cases}
\end{equation}
where $a$, $R_{c}$ and $Z$ are the softness parameter, core radius
and the effective s-electron occupancy, respectively. The term 
inside the core of Eq. (2) is deduced from the $d$-band scattering phase shift by using an inverse scattering 
approach. The term outside the core is just the Coulomb interaction between a
positive ion and a conduction electron. The coefficients of expansion in the core are related to the model
parameters as 
\begin{equation}
B_{1} = \frac{Z}{R_{c}}\left(1-\frac{2a}{R_{C}}\right)\,\exp\left(\frac{R_{C}}{a}\right) 
\end{equation}
\begin{equation}
B_{2} =  \frac{2Z}{R_{c}}\left(\frac{a}{R_{C}}-1\right)\,\exp\left(\frac{R_{C}}{a}\right)
\end{equation}

The effective partial pair potentials may be written within the framework of the pseudopotential theory as \cite{52}
\begin{equation}
 v_{ij}(r)=\frac{Z_{i}Z_{j}}{r}\left[1-\frac{2}{\pi}\int dq\,F_{ij}^{N}\, \frac{\sin(qr)}{q}\right]
\end{equation}
where, the indices $i$ and $j$ represent the ionic species $i$ and $j$.

In the above equaton, $F_{ij}^{N}$ is the normalized 
energy wave number characteristics and can be expressed as
\begin{equation}
 F_{ij}^{N}=\left[\frac {q^{2}}{\pi a \rho \sqrt{(Z_{i}Z_{j})}}\right]^{2}V_{i}(q)V_{j}(q)\left[1-\frac{1}
{\varepsilon(q)}\right]\left[\frac{1}{1-G(q)}\right]
\end{equation}
where, $V_{i}(q)$ denotes the local pseudopotential for the $i$th component and $\rho$  
the number density of ions, $\varepsilon(q)$ and $G(q)$ the dielectric 
screening function and the local field factor,  respectively.
Here, the dielectric function
\begin{equation}
\varepsilon(q)=1-\frac{\frac{4\pi e^{2}}{q^{2}}\chi(q)}{1+\frac{4\pi e^{2}}{q^{2}}G(q)\chi(q)}
\end{equation}
where $\chi(q)$ is the Lindhard function,

\begin{equation}
\chi(q)=-\frac{m k_{F}}{\pi^{2}\hbar^{2}}
\left[\frac{1}{2} +\frac{4k_{F}^{2}-q^{2}}{8qk_{F}}\ln \left| \frac{2k_{F}+q}{2k_{F}-q} \right| \right]
\end{equation}
\subsection{The LWCA theory}
The starting point of the LWCA theory \cite{53} is the thermodynamic perturbation theory \cite{54}. The blip functon in the theory can 
be written as,
\begin{equation}
B(r)=Y_{\sigma}\left\{\exp[-\beta v(r)]- \exp[\beta v_{\sigma}(r)]\right\}
\end{equation}
hence,
$v(r)$ and $v_{\sigma}(r)$ are the soft and hard sphere (HS) potentials, respectively. ${\beta}$ is the inverse 
temperature divided by the Boltzmann constant, and $Y_{\sigma}(r)$ known as the cavity function. $Y_{\sigma}(r)$
is associated with the hard sphere distribution function and is continuous at $r = \sigma$.
If we plot $B(r)$, as a function of r, it will give two sharp 
teeth shaped feature. In the LWCA theory this has been approximated
by right angle triangles. The Fourier transformation of B(r) at $k=0$ is then 
expanded in terms of Bessel's functions. Thus the thermodynamic condition
that for an effective hard sphere diameter $B(k=0)$  vanishes, leads to the transcendental 
equation,
\begin{equation}
 \beta v(\sigma)= \ln \left[ \frac{-2 \beta \sigma v'(\sigma) + Y +2}
{-\beta \sigma v'(\sigma) + Y +2} \right]
\end{equation}
for elemental system \cite{25}. In the case of alloys $v$ is replaced by $v_{ii}$ and $\sigma$ by $\sigma_{ii}$.
Equation (10) is solved graphically to obtain the effective hard sphere diameter. We note here
that, $\sigma_{12}$ determined from equation (10) is not strictly additive, but yields a value
very close to ($\sigma_{11}+\sigma_{22}$)/2 (deviation is  about 0.1\% only). So, following Ref. 55 we use 
the average value ($\sigma_{11}+\sigma_{22}$)/2 for $\sigma_{12}$ in order for compliance with the additive
hard sphere theory used in the present study.
\subsection{Pair distribution function}
We calculate the Ashcroft-Langreth (AL) partial structure factors, $S_{ij}(q)$,
in line with their original work  \cite{56}. In order to evaluate partial structure 
factors, the essential ingredients are the concentrations of two spheres in the
mixture and the effective HSD. The values for the effective HSDs for binary alloy
are obtained from the LWCA theory and are presented in Table I. The partial 
pair correlation functions necessary for real space calculation are derived from the 
Fourier transformation of S$_{ij}(q)$ in the following way:
\begin{equation}
g_{ij}(r)= 1+ \frac{1}{(2\pi)^{3}\rho \sqrt{C_{i}C_{j}} }
\int_{-\infty}^{\infty}(g(r)-1)\,e^{i\vec{q}.\vec{r}}\,d^{3}r
\end{equation}
where 
${\rho}$ is the ionic number density of the alloy and C$_{i}$ denotes the 
concentration of the $i$-th component. 
\begin{figure}
\begin{center}
\includegraphics[width=7cm,height=7cm,angle=270]{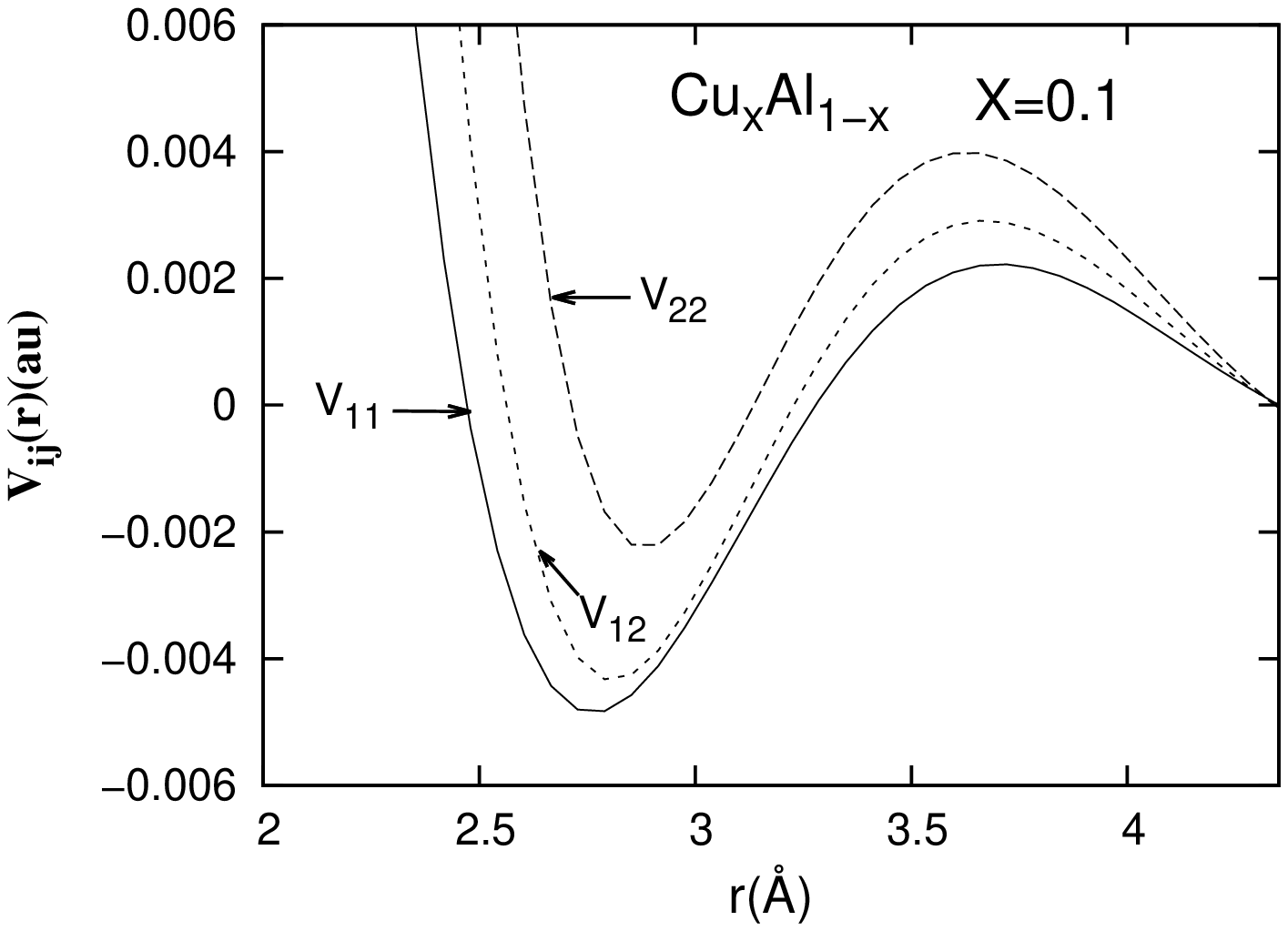}
\includegraphics[width=7cm,height=7cm,angle=270]{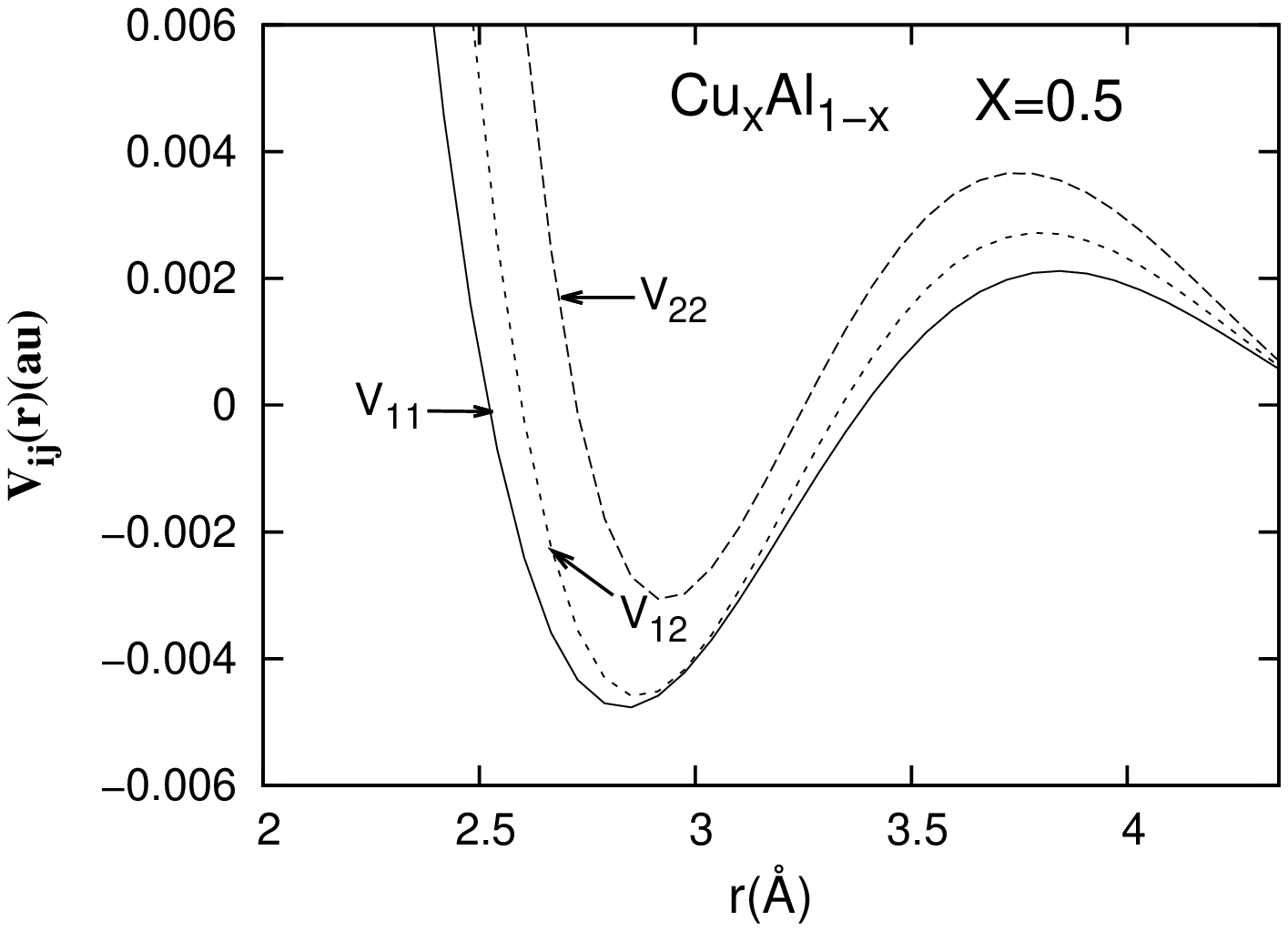}
\includegraphics[width=7cm,height=7cm,angle=270]{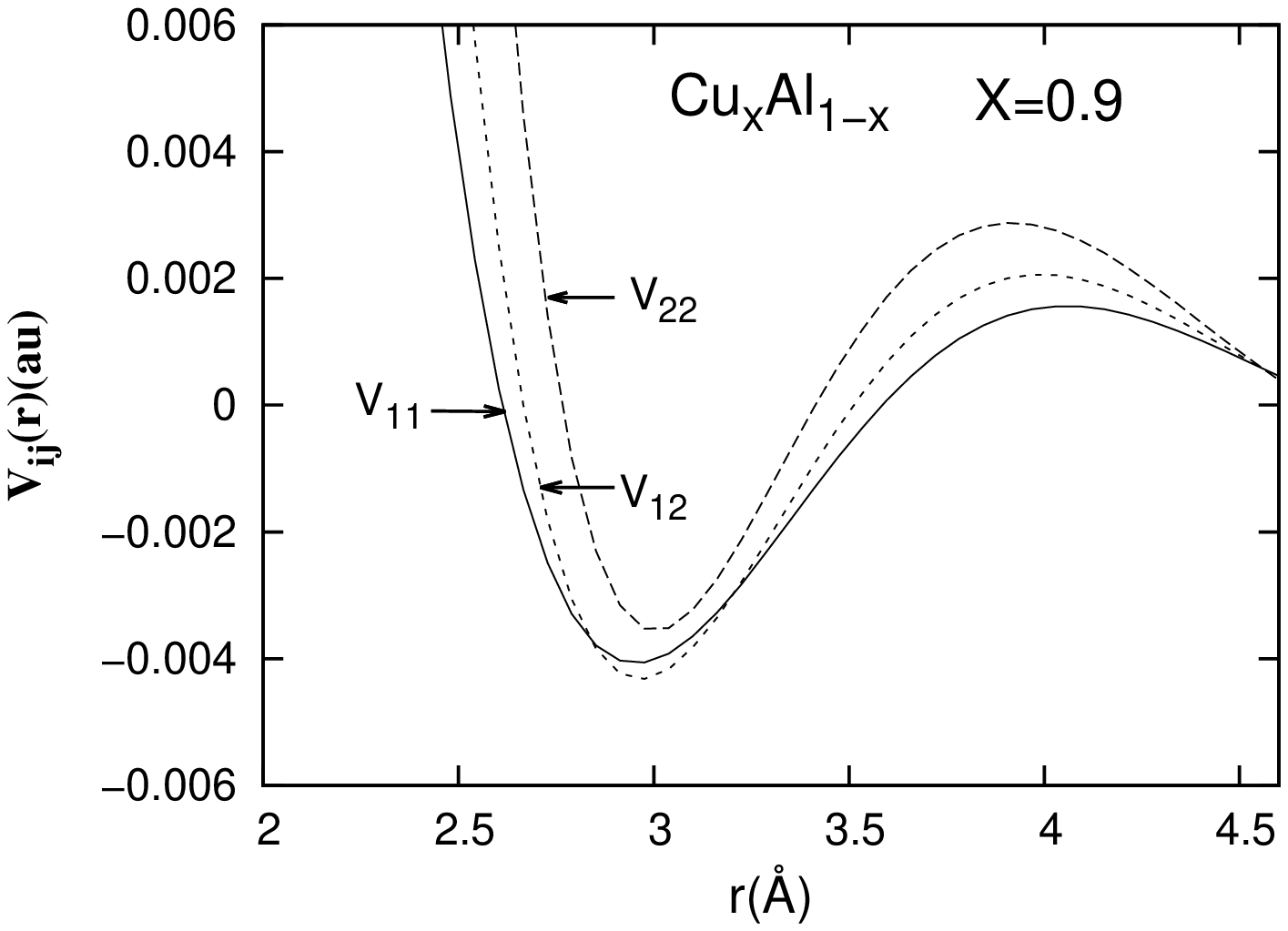}
\end{center}
\caption{Partial pair potentials for Cu$_{x}$Al$_{1-x}$ liquid binary alloys 
         for concentrations x=0.1, 0.5, 0.9, respectively.} 
\end{figure}
\subsection{Energy of mixing for liquid binary alloys}
Within the first order perturbation theory, the Helmholtz free energy per ion for an alloy 
may be written, in general as,
\begin{equation}
 F = F_{vol}+F_{eg}+F_{HS}+F_{Tail}
\end{equation}
where subscripts vol, eg, HS, Tail denote volume dependent, electron gas, 
hard sphere and tail of potential contributions, respectively.  
The volume contribution term can be written as \cite{57},
\begin{equation}
F_{vol} = \frac{1}{32\pi^{3}}
\int_{0}^{\infty}{q^{4}}\left\{{\frac{1}{\varepsilon(q)}-1}\right\}|\,v_{i}(q)|^{2}dq
-\frac{ZE_{F}}{3P}
\end{equation}
where, 
$E_{F}$  is the Fermi energy for electrons. And Z is the effective valence of alloy, 
$(Z= xZ_{1}+(1-x)Z_{1-x})$, 
$v(q)$ is the average  form factor of the electron-ion interaction for the alloy,
${\varepsilon(q)}$ is the dielectric function. Here, in Eq. (13) $P=\chi_{el}/\chi_{F}$, where $\chi$
denotes the isothermal compressibility and subcripts $el$ and $F$ stand for interacting and free electron. 

Now, the electron gas contribution to the free energy per electron may be written
as (\mbox{in Rydbergs units})
\begin{equation}
\frac{F_{eg}}{Z} = \left[\frac{2.21}{{r_{s}}^{2}} - \frac{0.916}{{r_{s}}}
 + 0.031\ln r_{s} - 0.115 \right]  
\end{equation}
where, $r_{s}$ is the dimensionless parameter defined as
\begin{equation}
 r_{s} = {(\frac{3}{4 \pi \rho Z })^{\frac {1}{3}}}/a_{0}
\end{equation}
 $a_{0}$ being the first Bohr radius. The ionic number density of the alloy

\begin{equation}
 \rho=\frac{\rho_{1} \rho_{2}}{(C_{1} \rho_{2} +C_{2} \rho_{1})}.
\end{equation}
Free  energy  per  atom  of  the  reference  HS  liquid  (using
$C_{1} = x$ and $C_{2} =1-x)$ is,
\begin{eqnarray}
\frac{F_{HS}}{k_{B}T} &=& \sum_{i}\left[-\ln(\varLambda_{i}^{3}v)+\ln 
 C_{i}\right] \nonumber\\
 &-& \frac{3}{2}\left(\frac{5}{3}-y_{1} +y_{2}+y_{3}\right) \nonumber\\
 &+&\frac{(3 y_{2}-2 y_{3})}{(1-\eta)} \nonumber\\
 &+&\frac{3}{2} \frac {(1-y_{1}+y_{2}+\frac {y_{3}}{3})}{(1-\eta)^{2}} \nonumber\\
 &+&(y_{3}-1) \ln(1-\eta),
\end{eqnarray}
where,
\begin{equation}
\varLambda_{i}=\left(\frac{2\pi\hbar^{2}}{{{m_{1}}^{C_{1}}}{m_{2}}^{C_{2}} k_{B} T }\right)^{\frac{1}{2}},
\end{equation}
 $m_{i}$ is the ionic mass and, packing fraction is 
\begin{align}
\eta=\sum_{i}\eta_{i}; \nonumber \\
\eta_{i} = \frac{C_{i}\pi\rho_{i}\sigma_{ii}^{3}}{6}
\end{align}
here, $C_{i}$ is the atomic concentration of $i$-th component,
\begin{equation}
 y_{1}=\sum_{j>i}\Delta_{ij}(\sigma_{ii} + \sigma_{jj})/{(\sigma_{ii}\sigma_{jj})}^{\frac{1}{2}}
\end{equation}
and, $\sigma_{ii}$ and  $\sigma_{jj}$ are  the additive hard sphere diameter (HSD).
\begin{equation}
 y_{2}=\sum_{j>i}\Delta_{ij}\sum_{k}(\frac{\eta_{k}}{\eta})^{\frac{1}{2}}/\sigma_{kk}
\end{equation}
\begin{equation}
y_{3} = \left[\sum_{i}(\frac{\eta_{i}}{\eta})^{\frac{2}{3}}C_{i}^{\frac{1}{3}}\right]^{3} 
\end{equation}

\begin{equation}
\Delta_{ij}= \left[(\eta_{i}\eta_{j})^{\frac{1}{2}}/\eta\right] \left[(\sigma_{ii}-\sigma_{jj})^{2}/\sigma_{ii}\sigma_{jj}\right](C_{i}C_{j})^{\frac{1}{2}}.
\end{equation}

In the above equation $C_{i}, \rho_{i}$, and $\sigma_{ii}$ denote the atomic concentration, 
ionic number density, and the effective hard sphere diameter (HSD) of the $i$-th component,
respectively. The contribution of tail part is written as,
\begin{equation}
 F_{Tail}=D\sum_{ij}C_{i}C_{j}M_{ij}
\end{equation}
where, $C_{i}$ and $C_{j}$ is the atomic concentration of $i$-{th} and $j$-{th} component,
respectively, and
\begin{equation}
  D=2 \pi \rho     
\end{equation}
$\rho$ is the ionic number density and
\begin{equation}
 M_{ij}=\int_{\sigma}^{\infty}v_{ij}(r)g_{ij}^{HS}(r,\sigma_{ij},\rho)\,r^{2}dr
\end{equation}
where, $v_{ij}$(r)  and $g_{ij}$(r)  are partial  pair potential  and pair
correlation  functions,  respectively.

Finally, the energy of mixing for Cu$_{x}$Al$_{1-x}$ liquid binary alloy reads,
\begin{eqnarray}
\Delta F&=&  F -\sum_{i}C_{i}F^{(i)}  \nonumber \\
&=& \Delta F_{vol}+\Delta F_{HS}+\Delta F_{eg}+\Delta F_{Tail}
\end{eqnarray}
where, $F^{(i)}$ represents the free energy for the $i$-th elemental component.

\begin{table}
\caption{\label{tbl-1} Partial hard sphere diameters for Cu$_{x}$Al$_{1-x}$ liquid binary
                 alloys for different concentrations at $T=1373$ K.}
\begin{tabular}{p{2.30cm}  p{2.30cm} p{0.90cm}}
\\
\hline
$x$ & $\sigma_{11}(\AA)$ & $\sigma_{22}(\AA)$  \\
\hline
0.0     & 0.00    & 2.6924  \\
0.1     & 2.480   & 2.6846   \\ 
0.2     & 2.489   & 2.6912    \\ 
0.3     & 2.499   & 2.6963     \\ 
0.4     & 2.508   & 2.7004      \\ 
0.5     & 2.516   & 2.7034       \\ 
0.6     & 2.522   & 2.7034        \\ 
0.7     & 2.527   & 2.7032         \\ 
0.8     & 2.531   & 2.6968          \\ 
0.9     & 2.532   & 2.6780           \\ 
1.0     & 2.547   & 0.00              \\
\hline 
\end{tabular} 
\end{table}

\begin{figure}
\begin{center}
\includegraphics[width=7cm,height=7cm,angle=270]{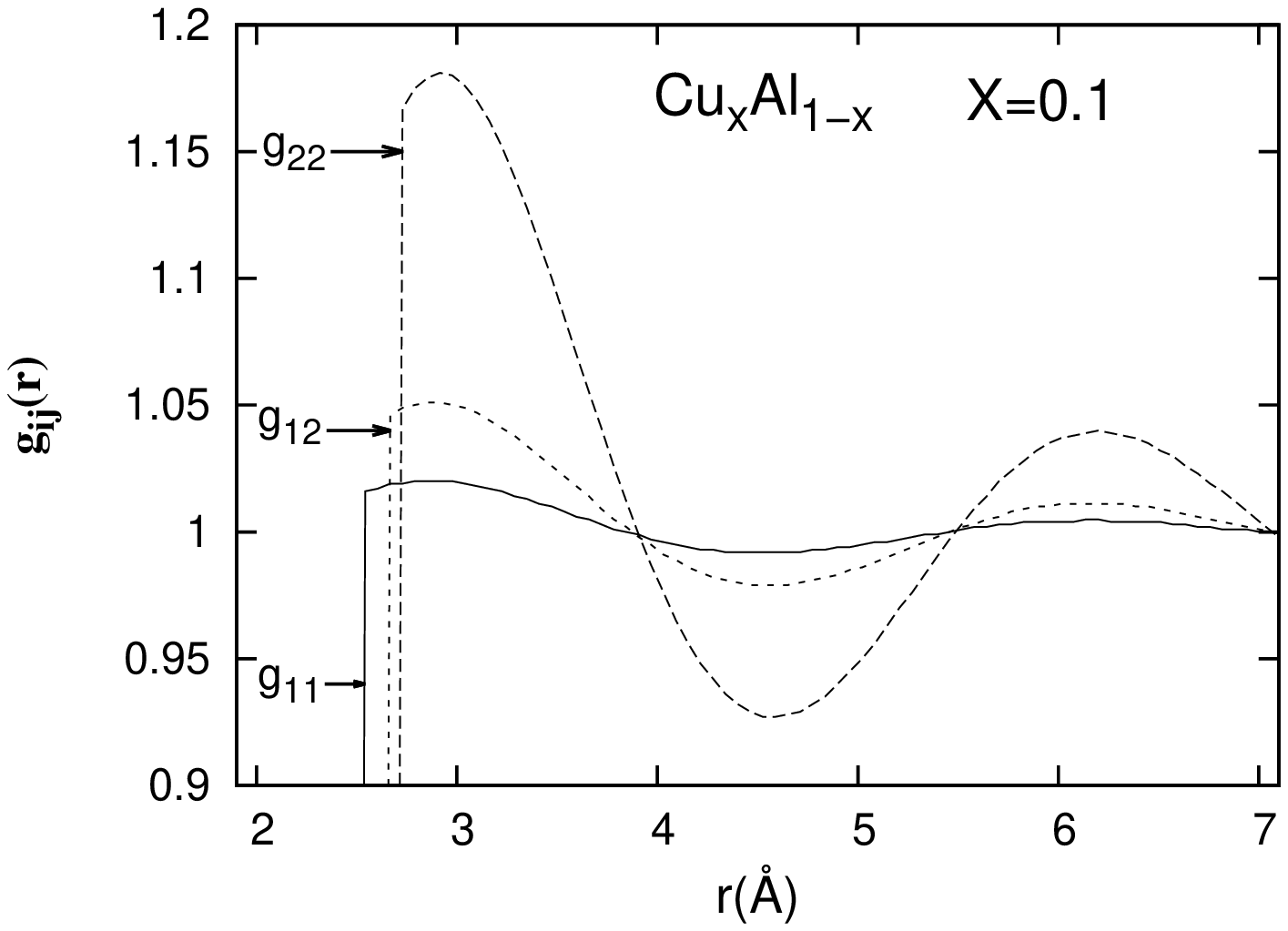}
\includegraphics[width=7cm,height=7cm,angle=270]{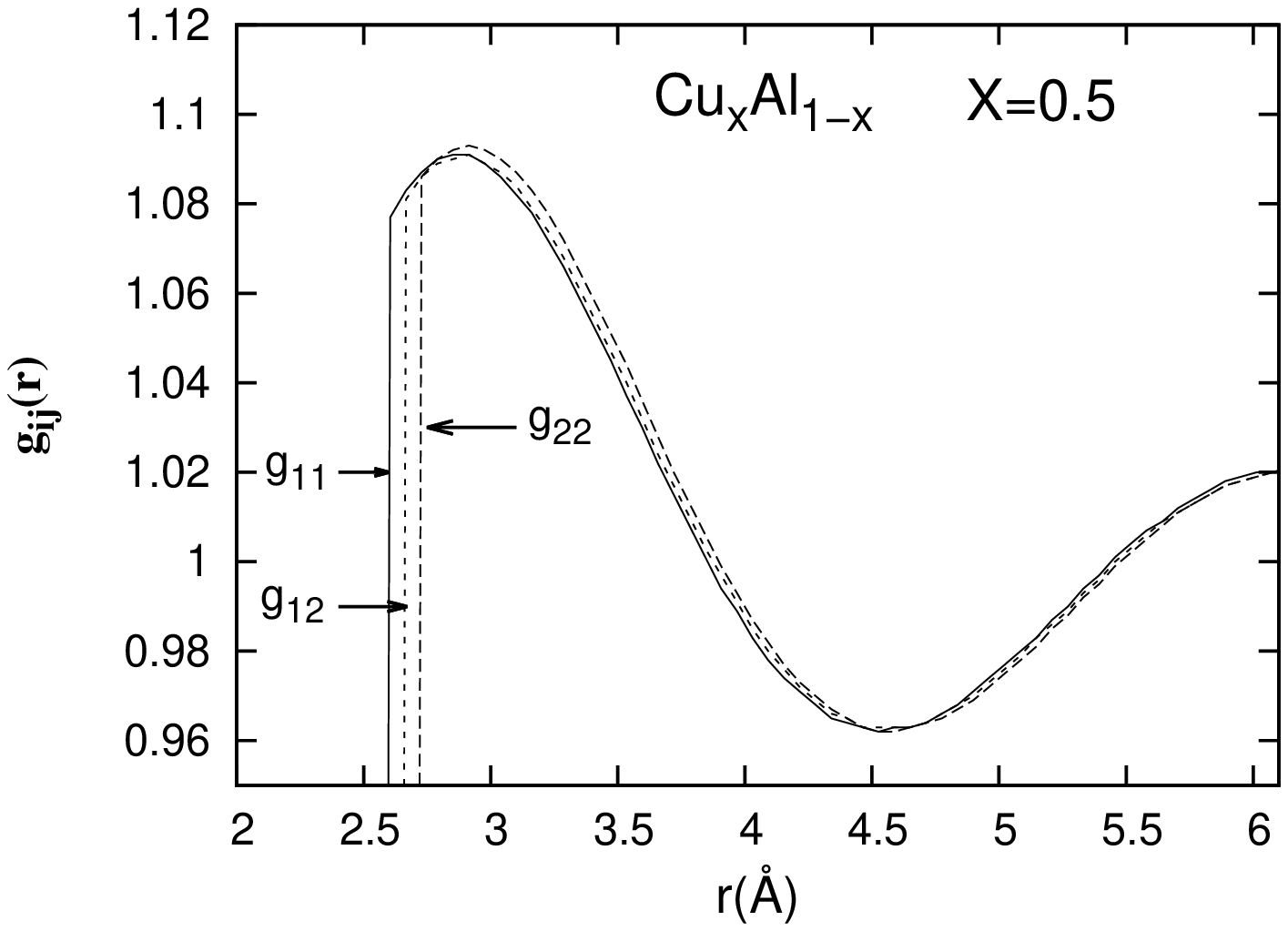}
\includegraphics[width=7cm,height=7cm,angle=270]{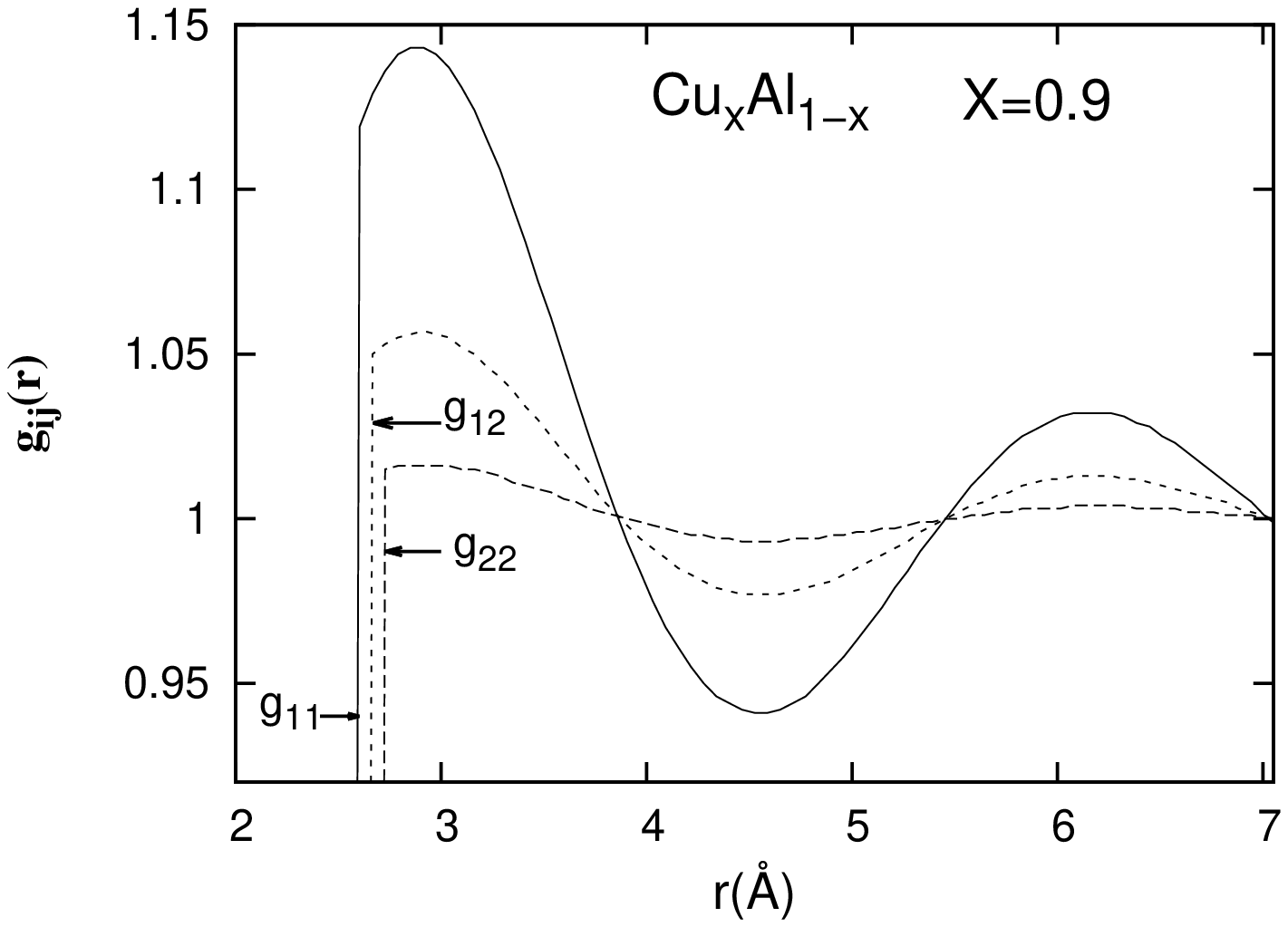}
\end{center}
\caption{Pair correlation functions for Cu$_{x}$Al$_{1-x}$ liquid binary alloys 
        for concentrations  x=0.1, 0.5, and 0.9, respectively.} 
\label{ent_alloy}
\end{figure}
\subsection{Entropy of mixing}                          

In the the perturbative approach the entropy per ion in the unit of
$k_{B}$ may be expressed for the liquid binary alloy as \cite{35}
\begin{equation}
\frac{S}{Nk_{B}}=\frac{S_{HS}}{Nk_{B}}+\frac{S_{Tail}}{Nk_{B}}
\end{equation}
where $S_{Tail}$ is the contribution of the tail of the interionic
interaction and, the hard sphere contribution $S_{HS}$ is given by
\begin{eqnarray}
\frac{S_{HS}}{Nk_{B}}=\frac{S_{ideal}}{Nk_{B}}+\frac{S_{gas}}{Nk_{B}}+\frac{S_{\eta}}{Nk_{B}} 
+ \frac{S_{\sigma}}{Nk_{B}}
\end{eqnarray}
The ideal term is written as
\begin{equation}
\frac{S_{ideal}}{Nk_{B}}=-\left[x \ln x+(1-x)\ln (1-x)\right]
\end{equation}
where $x$ is the concentration of the first element.
The gas term,
\begin{equation}
\frac{S_{gas}}{Nk_{B}}=\frac{5}{2}+\ln \left[\frac{1}
{\rho}\left\{ \frac{m_{1}^{x}m_{2}^{(1-x)}k_{B}T}{2\pi\hbar^{2}}\right\} ^{3/2}\right],
\end{equation}
the packing term,
\begin{equation}
\frac{S_{\eta}}{Nk_{B}}=\ln(1-\eta)+\frac{3}{2}[1-(1-\eta)^{-2}],
\end{equation}
and the diameter mismatch term,
\begin{eqnarray*}
\frac{S_{\sigma}}{Nk_{B}}=\Biggl[\frac{\pi x\,(1-x)\rho \,(\sigma_{11}-\sigma_{22})^{2}\,(1-\eta)^{-2}}{24}\Biggr]
\end{eqnarray*}
\begin{eqnarray}
 & \times\Biggl\{\Bigl[12\,(\sigma_{11}+\sigma_{22})
-\pi\rho \,(x\sigma_{11}^{4}+(1-x))\sigma_{22}^{4})\Bigr]\Biggr\},
\end{eqnarray}
$\eta$ denotes the packing fraction 
\begin{equation}
\eta=\frac{\pi\rho}{6}\bigl[x \sigma_{11}^{3}+(1-x))\sigma_{22}^{3}\bigr].
\end{equation}

The tail part of the entropy for the elemental system may be expressed
as 
\begin{equation}
\frac{S_{Tail}}{Nk_{B}}=\frac{1}{Nk_{B}}\Biggl(\frac{\partial F_{Tail}}{\partial T}\Biggr)_{\Omega},
\end{equation}
and for the binary alloys as 
\begin{eqnarray}
\frac{S_{Tail}^{Alloy}}{Nk_{B}}&=&\frac{1}{Nk_{B}}\Biggl[\biggl
(\frac{\partial F_{Tail}}{\partial T}\biggr)_{\Omega,\rho,
\sigma_{ii}} \nonumber\\
&+& \sum_{i=1}^{2}\biggl(\frac{\partial F_{Tail}}{\partial
\sigma_{ii}}\biggr)_{\Omega,T}\biggl(\frac{\partial\sigma_{ii}}{\partial T}\biggr)_{\Omega,\rho}\Biggr].
\end{eqnarray}
The values for $(\partial F_{Tail}/\partial T)_{\Omega,\rho,\sigma_{ii}}$ and
$(\partial F_{Tail}/\partial\sigma_{ii})_{\Omega,T}$ are solved numerically
by the process of partial differentiation.
%
To evaluate $\partial\sigma/\partial T$, we used the well known formula
due to protopapas $et$ $al$. \cite{58},
\begin{equation}
\sigma=1.126\sigma_{m}\biggl\{1-0.112\Bigl(\frac{T}{T_{m}}\Bigr)^{1/2}\biggr\}
\end{equation}
\begin{equation}
\frac{\partial\sigma_{i}}{\partial T}=-0.063\sigma_{im}\frac{1}{\left(T_{m}T\right)^{1/2}};
\end{equation}
where the subscript $m$ indicates the thermodynamic properties at melting temperature. 
This equation is adequate for a good qualitative description of the physical properties under study.

Finally, the formula for the entropy of mixing stands as  
\begin{equation}
\Delta S=S-\sum_{i}C_{i}Si=\Delta S_{HS}+\Delta S_{Tail}.
\end{equation}
\section{Results and Discussion}
The results of claculations for energy of mixing and entropy of mixing are presented in this section for 
Cu$_{x}$Al$_{1-x}$  liquid binary alloys. The perturbation theory along 
with the electronic theory of metal is applied to perform the calculations. The basic 
ingradients required for the real space calculation are the effective interionic partial
potentials $V_{ij}(r)$ and the corresponding pair correlations functions of the reference
system.

Elements forming $Cu_{x}Al_{1-x}$  alloys are Al and Cu; the first element is a polyvalenet simple metal
and the second one is a noble metal often recognized as the $3d$ transition  metals,
because of the presence of the $sp$-$d$ hybridization effect. So, we have employed the BS
model to describe the interionic interaction of both the simple and transition metals 
simultaneously. Three parameters of the BS model are the core radius, $R_{c}$, the softness 
parameter, $a$, and the valence $Z$. The value of $R_{c}$ is generally fixed by fitting 
physical properties of the concerned systems \cite{59,60}. We have taken the values of 
$R_{c}$ for $l$-Al from Ref. 61 and for l-Cu from Ref. 62. The value
for the former is 1.91 $au$ and for the latter is 1.44 $au$. The softness parameter '$a$' 
is determined by the best fitting of the LWCA structure factor to the experimental ones 
\cite{47}. The values of '$a$' thus found for Al and Cu are 0.49 and 0.30 $au$, rspectively.
The quality of fit is shown in the Appendix. Regarding the third parameter called effective 
$s$-electron occupancy number, $Z_{s}$, we take the chemical valenece $3$ for Al and, 1.32 for Cu due to 
the presence of $sp$-$d$ hybridization effect in Cu \cite{62}. We note here that, the dielectric function
(see Eqn. (7)) plays an important role in determining the effective pair potential profile. In this work we
have used the dielectric function proposed by Icimaru and Utsumi \cite{63}, because, their theory satisfies the 
compressibility sum rule and the short range correlation conditions for the wide range of metallic density.   
\begin{figure}
\begin{center}
\includegraphics[width=7cm,height=7cm,angle=270]{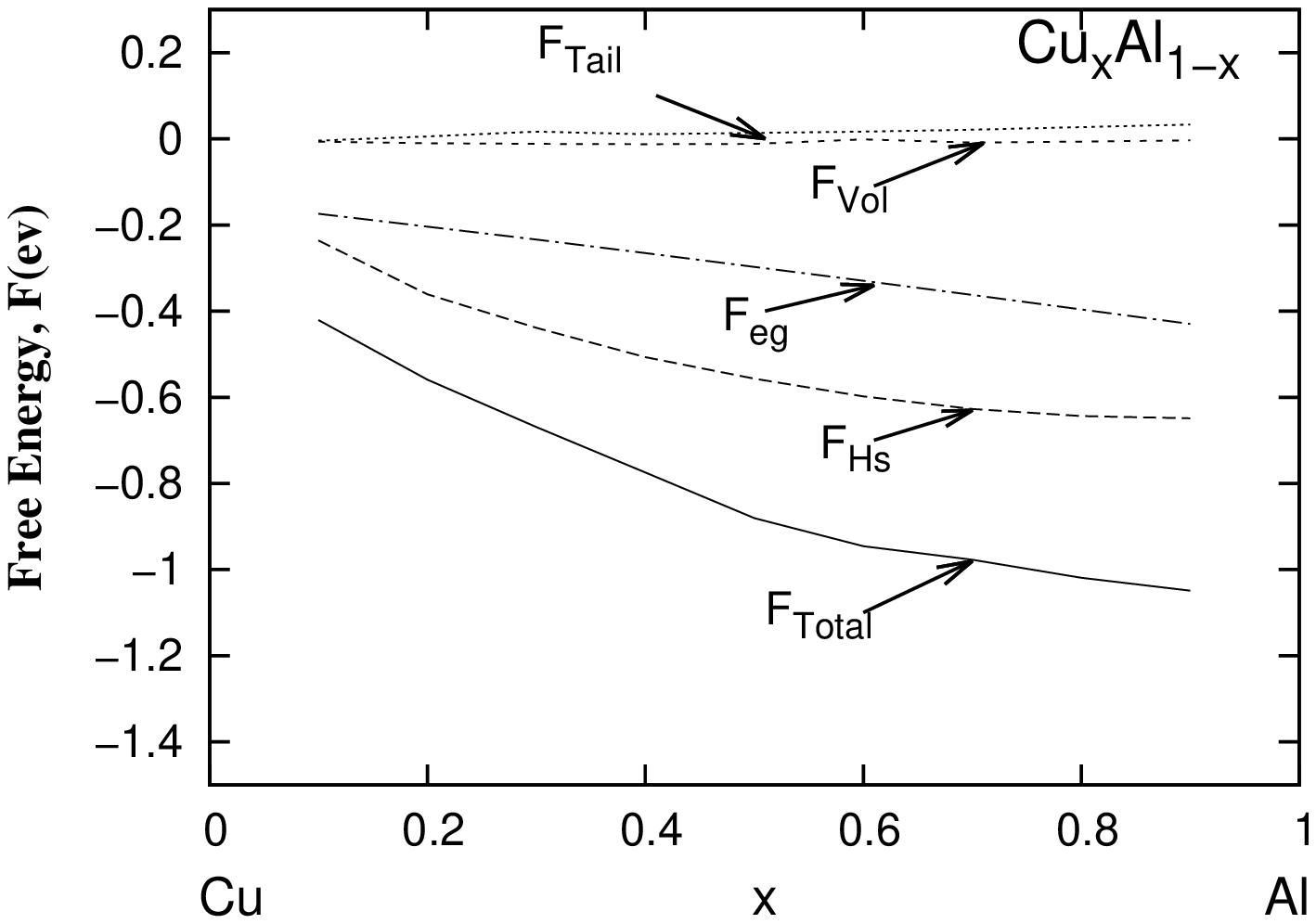}
\end{center}
\caption{Total Energyy as a function of concentration at T=$1373$ K for Cu$_{x}$Al$_{1-x}$ 
         liquid binary alloys.} 
\label{ent_alloy}
\end{figure}

FIG.1. shows the effective partial pair potentials for Cu$_{x}$Al$_{1-x}$ liquid binary alloys
for three different concentrations, specificly for x=0.1, 0.5 and 0.9. The position of the 
principal potential minima and the depth of the potential well are results of deliate balance between the 
repulsive and attractive interactions in metals. In the pseudopotential formalism this is accomplished
by the direct interaction between different ion cores and by the indirect interaction mediated
by conduction electrones. In this case the dielectric function plays an important role in determining
the appropriate potential profile. From FIG. 1 it is noticed that the depth of the potential well is
the largest for $V_{Cu-Cu}$ and the smallest for $V_{Al-Al}$. The well for $V_{Cu-Al}$ lies in between 
for concentration $x < 0.9$. For $x=0.9$ the potential well for $V_{Cu-Al}$ goes below to that of $V_{Cu-Cu}$. 
This feature is very unusual and we have not observed in any alloy we have studied so far. In all cases we
found $V_{12}$ in between $V_{11}$ and $V_{22}$ without any exception. It 
is known that the posititive magnitude of $V_{Ord}= V_{Cu-Al}- (V_{Cu-Cu}+V_{Al-Al})/2$ indicates
a tendency of segregation to occur and, the negative magnitude indicates a tendency of mixing. 
But if the negativity is unusally large, as in the present case, one may attribute it to the tendency of compound 
formation. But drawing a concret conclusion regarding this feature requires further study in detail. 
\begin{figure}
\begin{center}
\includegraphics[width=7cm,height=7cm,angle=270]{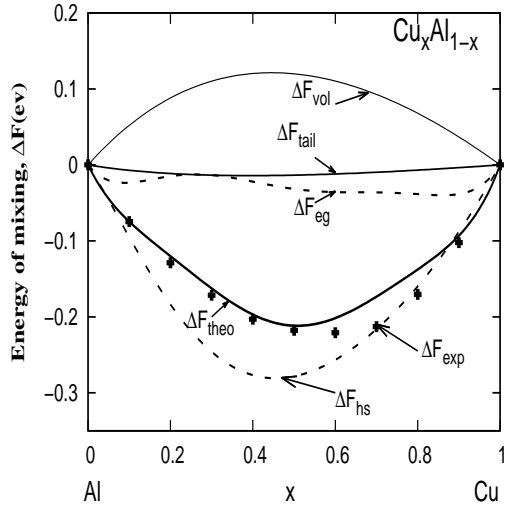}
\end{center}
\caption{Energy of mixing as a function of concentration at T=$1373$ K for Cu$_{x}$Al$_{1-x}$ 
         liquid binary alloys.} 
\label{ent_alloy}
\end{figure}

Partial pair correlation function $g_{ij}$ for liquid Cu$_{x}$Al$_{1-x}$ binary alloys calculated
from the LWCA theory are shown in FIG. 2. It is noticed that for $x=0.1$, i.e. in the Al rich alloys
the principal peak of $g_{Al-Al}(r)$ is much larger than that of $g_{Cu-Cu}(r)$ and this trend reverses 
for  $x=0.9$, that is when the alloy becomes rich in Cu concentration. The simple reason of it is, in the
Al rich alloys the probability of finding a second Al ion from the one at origin is higher than that of 
finding a Cu ion and vice versa. This trend in principle, suggests that for $x=0.5$ the principal peak
value of both $g_{Cu-Cu}(r)$ and  $g_{Al-Al}(r)$ should be of equal magnitudes. This is exactly reflected 
in FIG. 2(b). We note here that, for some alloys a slight variation of peak values of $g_{Cu-Cu}(r)$ and  $g_{Al-Al}(r)$
for $x=0.5$ might be arised due to size difference of hard sphere. TABLE-I illustrates the values
of effective hard sphere diametres determined from the LWCA theory. A close look at the table reveals that
the HSD $\sigma_{Cu-Cu}$ increases  and that of $\sigma_{Al-Al}$ decreases with increasin $Cu$ concentration. This
feature of variation of ionic size may be attributed to the charge transfer that occurs from one species
of atom to another species in the alloy state \cite{64}.
\subsection{Energy of mixing}
FIG. 3 illustrates  different parts of Helmholtz free energy for  Cu$_{x}$Al$_{1-x}$ liquid binary
alloy obtained using the ETM in conjunction with the perturbation theory. It is seen that the largest
contribution arises from the HS reference system. This actually menifests the principle of
perturbation approach in which the reference system mostly resembles the real system. FIG. 3 also shows
that the smallest contribution to the total free energy is coming out from the tail part of the 
interaction. The magnitude of the volume  dependent (i.e structure independent) component of the 
free energy, $F_{Vol}$, is very close but slightly lower than that of the $F_{Tail}$ contribution. The electron
gas part of the total free energy, $F_{eg}$, lie in between  the $F_{Vol}$ and $F_{HS}$. Symbolically
one can present the magnitude of contribution as in the following order $F_{Tail}<F_{Vol}<F_{eg}<F_{HS}$. Finally the summation
of all four contributions yields the total free energy $F_{Total}$ of the alloy which lies at the bottom 
in the Figure.

FIG. 4 shows the breakdown details of the energy of mixing for Cu$_{x}$Al$_{1-x}$ liquid binary alloy calculated at temperature $T=1373$ K. It is seen that the largest contribution to the total energy of mixing arises from the hard sphere reference system and these are negetive across the full range of concentration. This trend is similar to those of other systems (Ag$_{x}$In$_{1-x}$, Ag$_{x}$Sn$_{1-x}$) studied before \cite{34}. The second largest contribution comes out from the volume part of the interaction and this contribution $(\Delta F_{vol})$ is seen to be  positive for whole range of concentration. This feature is unlike other miscible alloys \cite{34}, but like to those of segregating alloys \cite{38,39}. The third and the fourth largest contributions come from the electron gas and tail part of the interactions, respectively. The total energy of mixing  obatined from the algebric sum of the four components agree well with the corresponding experimental data \cite{65}. More specifically, the agreement is excellent upto equiatomic concentration i.e. 
upto $x=0.5$, and after that it is just good from the point of view of experimental uncertainty ($\pm 0.01$ev). Most importantly the minimum of $\Delta F$ is found to be near the equatomic concentration as suggested by the experiment \cite{65}.
\begin{figure}
\begin{center}
\includegraphics[width=7cm,height=7cm,angle=270]{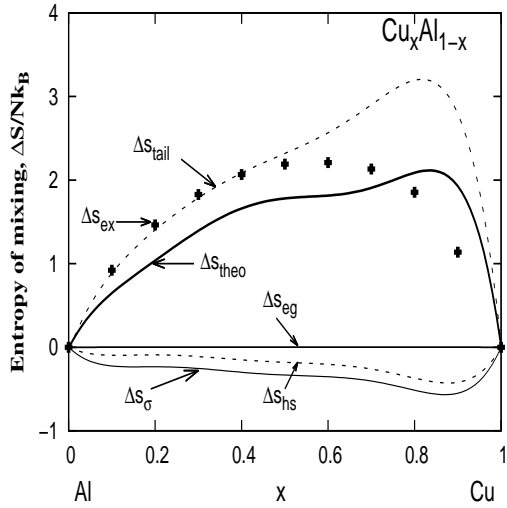}
\end{center}
\caption{Entropy of mixing as a function of concentration at T=$1373$ K for Cu$_{x}$Al$_{1-x}$ 
         liquid binary alloys.} 
\label{ent_alloy}
\end{figure}
\subsection{Entropy of mixing}
Equation (28) shows that both the hard and soft part of the interionic intaraction contribute to the total entropy of the system. The hard sphere part, $S_{HS}$, combines the ideal term $S_{id}$, the gas term, $S_{gas}$, packing term, $S_{\eta}$, and the HSD mismatch term, $S_{\sigma}$, together. The softpart arises  due to the attractive interaction represented by the tail of the potentials. There are evidences \cite{51} that the HS system alone cannot account for vibrational contribution to the entropy. We must note here that the contribution of the second term of equation (36) disappears if variational perturbation method is used instead of the perturbation approach without variation; thus the theory turns nearly into that of the HS alone. We believe that the perturbative approach alone would give in some way a contribution to compensate the vibrational effect in liquid alloys of our concerned alloys. The work by Kumaravadivel and Evans \cite{51}
lends support to our interpretation.
The breakdown details of the entropy of mixing, $\Delta S$, for liquid Cu$_{x}$Al$_{1-x}$  binary alloys at T=$1373$ K are shown in FIG. 5. It is seen that the tail part contribution to the entropy of mixing, $\Delta S_{tail}$, plays the dominant role in determining the total entropy of mixing. It is positive across the whole concentration range. The second largest contribution is arising from the HSD mismatch term, but it is negative across the whole range of concentration. The third largest contribution to $\Delta S$ comes from the HS reference system and it is denoted by $\Delta S_{HS}$. The contribution of the volume part of the interionic interaction, $\Delta S_{eg}$ is the least in the present case, the magnitude of it is almost zero. This feature of $\Delta S_{eg}$ is unlike that of the  $\Delta F_{eg}$, because the latter contribution to the energy of mixing is significant. FIG. 5 also shows that $\Delta S$ is not symmetric across the equiatomic concentration, it is rather assymmetric in nature. The maximum of $\Delta S$ is around concentration $x=0.8$, whereas the maximum of experimental data is at $x=0.6$. The discrepancy between the  positions of the maxima of theoretical and experimental $\Delta S$ may be attributed to the corresponding difference between the theory and experiment of $\Delta F$ in the Cu-rich alloy (see FIG.4). However, in the present study the overall agreement of $\Delta S$ with the experimental ones for whole range of concentration is not bad when the difficulty of calculation of entropy from the first principles method is cinsidered.
\begin{figure}
\begin{center}
\includegraphics[width=7cm,height=7cm,angle=270]{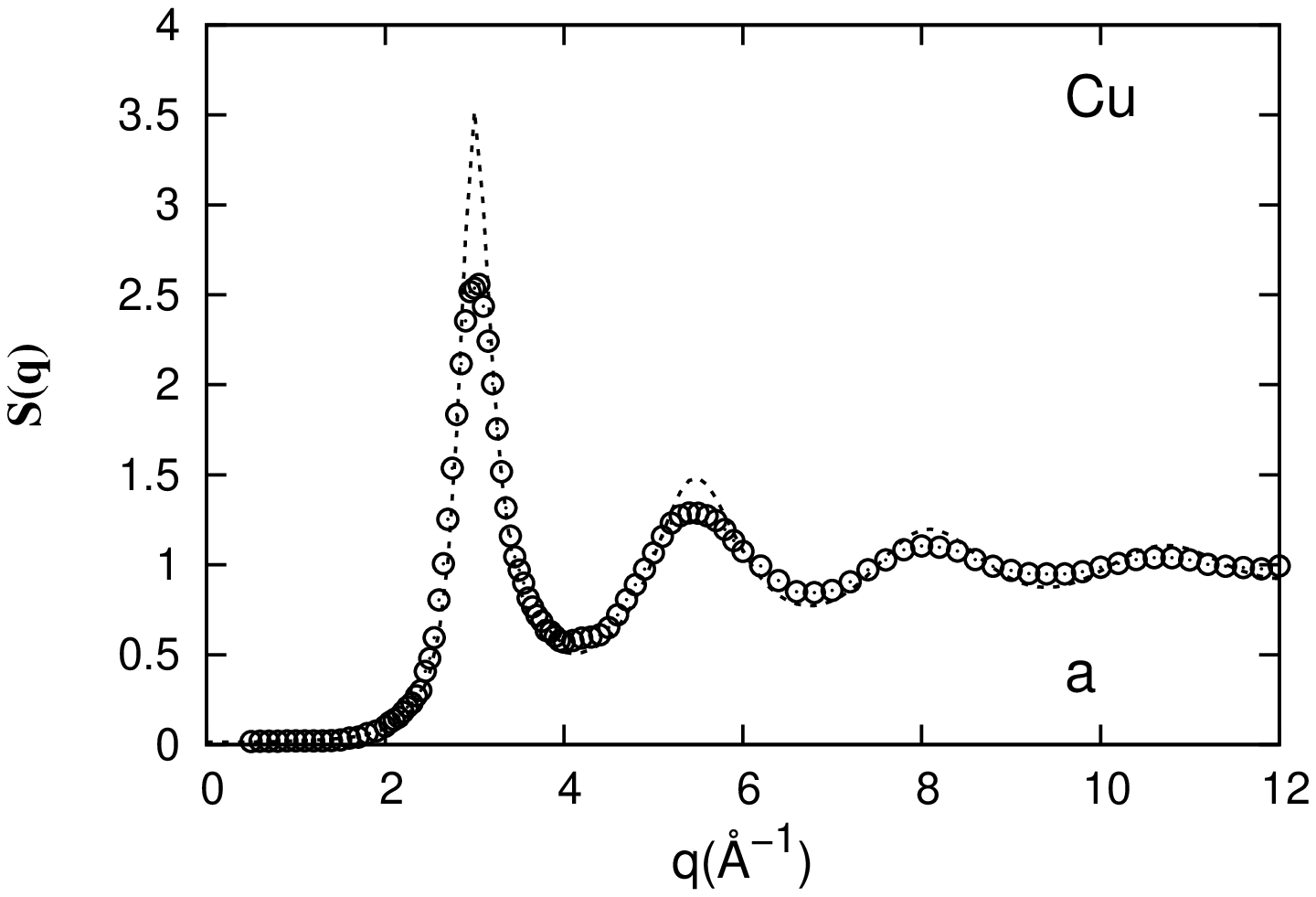}
\includegraphics[width=7cm,height=7cm,angle=270]{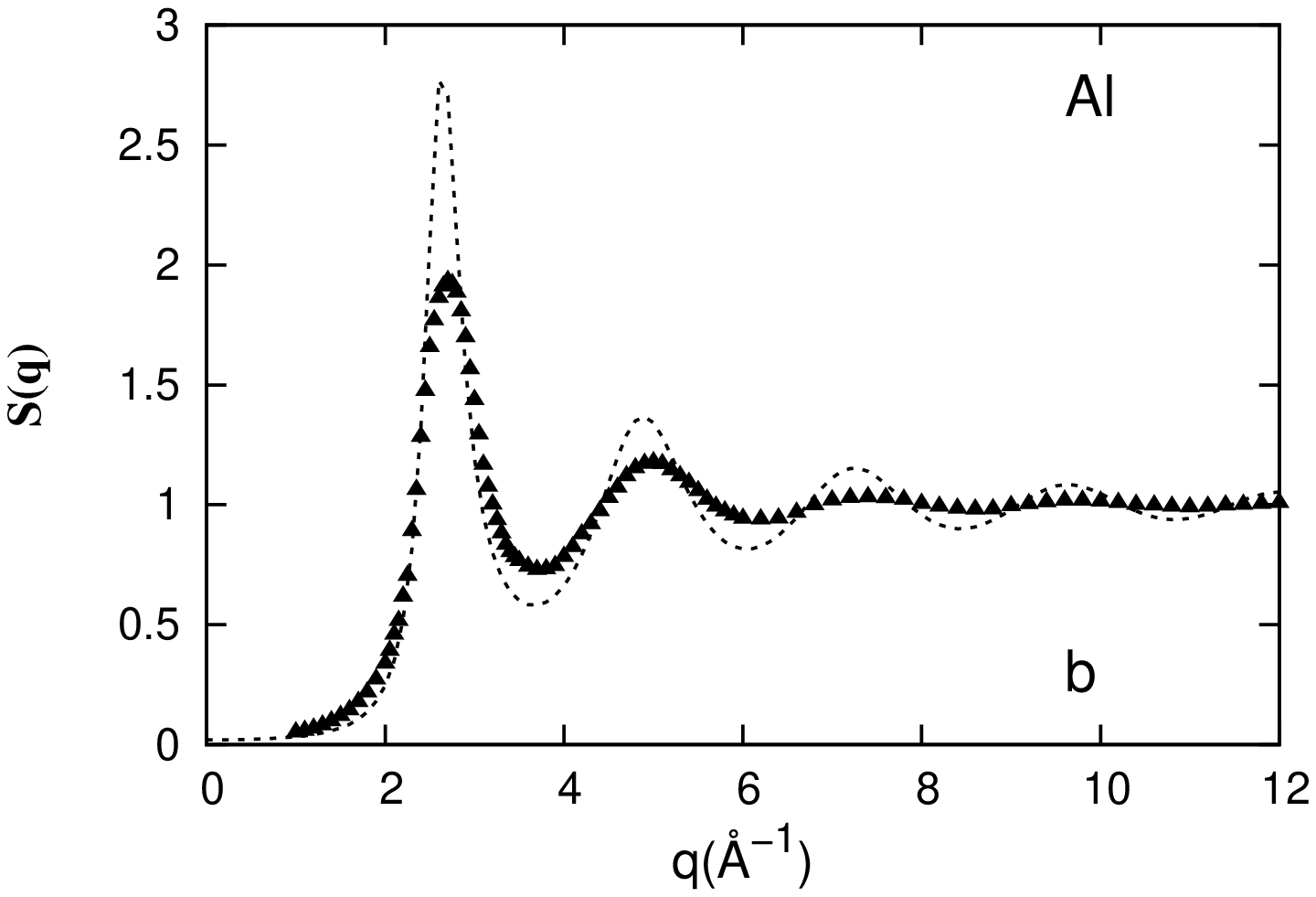}
\end{center}
\caption{Static structure factors for liquid Cu and Al. Triangles and rectangular bars represent the theoretical and experimental results, respectively.} 
\label{ent_alloy}
\end{figure}
\section{Conclusion}
We, in the present study, have systematically investigated the energy of mixing and the entropy of mixing for
liquid Cu$_{x}$Al$_{1-x}$ binary alloys employing the perturbation method along with the electronic theory of
metals. Although the Helmholtz free energy and entropy are closely related thermodynamically, the entropy of
mixing is much more difficult to calculate as mentioned before. This is because, the accuracy of the latter 
depends on the presise shape of the former one in $F$ versus $T$ curve which is difficult to have in numetical
calculation. From the above results and discussion we can, however, draw
the following conclusions. The ETM as described via a local pseudopotential (BS) model in conjunction with the 
perturbative  approach is able to describe the energy of mixing for Cu$_{x}$Al$_{1-x}$ binary alloys 
with a great degree of accuracy. In the case of entropy of mixing the calculation is completely
free from any adjustable parameter. From this point of view result for the entropy of mixing is
fairly good qualitatively. The main cause of discrepancy between theory and experiment  in the Cu-rich alloy is,
in our view, due to the existance of the complicated $d$-band characteristics including $s-d$ hybridization 
effect in Cu. In addition, the tendency to form a compound in the Cu rich alloy might also be responsible
to widen the discrepancy; in order to establish it further research is required. So, apprantly, a quantitative description of thermodynamic properties of Cu$_{x}$Al$_{1-x}$ liquid 
binary alloys from the present approach requires a precise account for the $d$-band effects 
in the interionic interaction for Cu. As our present approach provides total entropy of the elemental and alloy
systems it may be extended to the study of atomic
transport properties through the universal scaling laws\cite{19,20}. Our present approach for free energy calculation 
may also be applied to the study of temperature dependent properties such as in theory of melting for metallic systems.

\section{APPENDIX: Determination of the softness parameter}
The values of the softness parameter of the Bretonnet Silbert local potentials are determined by fitting the LWCA
static structure factors to the corresponding X-ray diffraction data \cite{47}. The quality of fit are shown in 
FIG.6.

\end{document}